\documentclass[prd,preprint,showpacs]{revtex4}
\usepackage{graphicx}
\usepackage{latexsym}
\usepackage{amsmath}
\begin{document}
\title{Direct evidence of acceleration from distance modulus redshift graph}
\author{Yungui Gong}
\email{gongyg@cqupt.edu.cn}
\affiliation{College of Electronic Engineering, Chongqing
University of Posts and Telecommunications, Chongqing 400065,
China}
\affiliation{GCAP-CASPER, Physics Department, Baylor University,
Waco, TX 76798, USA}
\author{Anzhong Wang}
\email{anzhong_wang@baylor.edu}
\affiliation{GCAP-CASPER,
Physics Department, Baylor University, Waco, TX 76798, USA}
\author{Qiang Wu}
\email{qiang_wu@baylor.edu}
\affiliation{GCAP-CASPER,
Physics Department, Baylor University, Waco, TX 76798, USA}
\author{Yuan-Zhong Zhang}
\email{zyz@itp.ac.cn}
\affiliation{Institute of Theoretical Physics, Chinese Academy of Sciences,
P.O. Box 2735, Beijing 100080, China}
\begin{abstract}
The energy conditions give upper bounds on the luminosity distance. We apply these
upper bounds to the 192 essence supernova Ia data to show that the Universe
had experienced accelerated expansion. This conclusion is drawn directly from
the distance modulus-reshift graph. In addition to be a very simple method, this
method is also totally independent of any cosmological model. From the
degeneracy of the distance modulus at low redshift, we argue that the
choice of $w_0$ for probing the property of dark energy is misleading.
One explicit example is used to support this argument.

\end{abstract}
\preprint{astro-ph/0703583}
\maketitle

\section{Introduction}

Ever since the discovery of the accelerated expansion of the Universe by the
supernova (SN) Ia observations \cite{agr98}, many efforts have been made
to understand the mechanism of this accelerated expansion. Although
different observations pointed to the
existence of dark energy which has negative pressure and contributes
about 72\% of the matter content of the Universe
\cite{riess,astier,riess06, wmap3,sdss}, the
nature of dark energy is still a mystery to us. For a review of dark energy models,
one may refer to Ref. \cite{DE}.

Due to the lack of a satisfactory dark energy model, many parametric
and non-parametric model-independent methods were proposed
to study the property of dark energy and the geometry
of the Universe \cite{virey,sturner,gong06,gong07,astier01,huterer,weller,alam,gong04,gong05,lind,jbp,par1,par2,
par3,par4,par5,gong04a,wang05,jbp05, nesseris6a,berger,sahni06,lihong,yun06,evldh,saini,jbp06,nesseris05}.
In the reconstruction of the deceleration parameter $q(z)$, it was found
that the strongest evidence of acceleration happens at
redshift $z\sim 0.2$ \cite{virey,sturner,gong06,gong07}. The sweet spot of
the equation of state parameter $w(z)$ was found to be
around the redshift $z\sim 0.2-0.5$ \cite{gong07,astier01,huterer,weller,alam,gong04,gong05}.

The energy conditions were also used to study the expansion of the Universe in
\cite{visser,alcaniz,aasen}. The energy condition $\rho+3p\ge 0$ is equivalent to $q(z)\ge 0$,
and the energy condition $\rho+p\ge 0$ is equivalent to $\dot{H}\le 0$ for a flat
universe. These conditions give lower bounds on the Hubble parameter $H(z)$, and therefore
upper bounds on the luminosity distance. These bounds can be put in the
distance modulus-redshift graph to give direct model independent evidence of accelerated expansion.
On the other hand, to the lowest order, the luminosity distance $d_L(z)$ is independent
of any cosmological model. In the low $z$ region ($z\le 0.1$), $d_L(z)$ is degenerate.
So different dark energy models will give almost the same $d_L(z)$ in the low
$z$ region and the current value $w_0$ of $w(z)$ is not well constrained. That
is the main reason why the sweet spot is found to be around $z\sim 0.3$. In this paper,
we compare two dark energy models which differ only in the low $z$ region to
further explain the consequence of the degeneracy.

This paper is organized as follows. In section II, we apply the energy
conditions to the flat universe to show model independent evidence of accelerated
expansion. In section III, we use two dark energy models to argue as to why
the value of $w_0$ is not good for exploring the property of dark energy.
The energy conditions are applied to the non-flat universe in section IV.
In section V, we conclude the paper with some discussions.

\section{Distance modulus redshift graph}

The strong energy condition $\rho+3p\ge 0$ and $\rho+p\ge 0$ tells us that
\begin{equation}
\label{sec}
q(t)=-\ddot{a}/(aH^2)\ge 0,\quad \dot{H}-\frac{k}{a^2}\le 0
\end{equation}

\begin{figure}
\centering
\includegraphics[width=14cm]{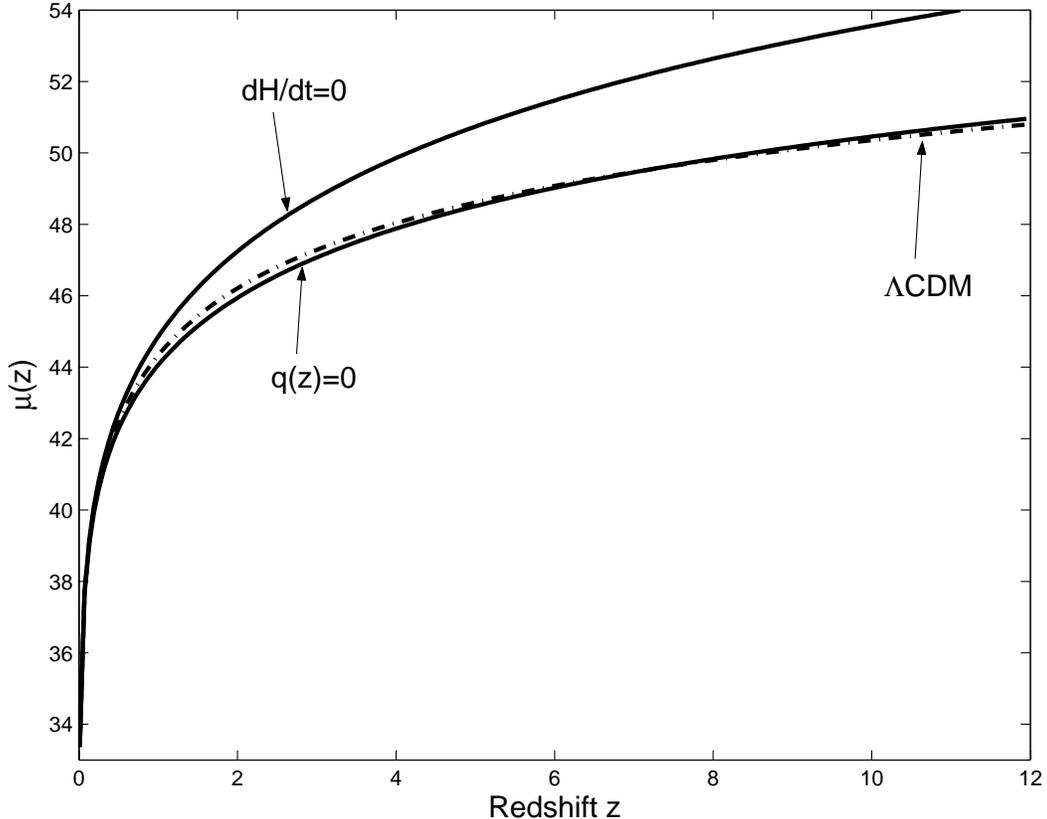}
\caption{The distance modulus $\mu(z)$. The  dash-dotted line corresponds to
the $\Lambda$CDM model with $\Omega_m=0.27$. The solid line corresponds to the
bound from the strong energy condition.}
\label{lcdm}
\end{figure}

The Hubble parameter $H(t)=\dot{a}/a$ and the deceleration
parameter $q(t)$ are related by the following equation,
\begin{equation}
\label{hubq}
H(z)=H_0\exp \left[\int^z_0 [1+q(u)]d\ln(1+u)\right],
\end{equation}
where the subscript 0 means the current value of the variable.
Therefore, the strong energy condition requires that
\begin{equation}
\label{sech}
H(z)\ge H_0(1+z),
\end{equation}
and
\begin{equation}
\label{sech1}
\quad H(z)\ge H_0\sqrt{1-\Omega_k+\Omega_k(1+z)^2},
\end{equation}
for redshift $z=a_0/a-1\ge 0$. Note that the satisfying of
Eq. (\ref{sech}) guarantees the satisfying of Eq. (\ref{sech1}).
Although Eq. (\ref{sech}) is derived from Eq. (\ref{sec}), they are not
equivalent. Due to the integration effect, we cannot derive Eq. (\ref{sec})
from Eq. (\ref{sech}).
If the strong energy condition is always satisfied, i.e., the Universe
always experiences decelerated expansion, then there is a lower bound
on the Hubble parameter given by Eq. (\ref{sech}).
When Eq. (\ref{sech}) is violated, we conclude that the Universe once experienced accelerated expansion.
On the other hand, if the Universe always experiences accelerated expansion,
then there is an upper bound on the Hubble parameter.
So if the Hubble parameter satisfies Eq. (\ref{sech}), then we conclude that the
Universe once experienced decelerated expansion.
The satisfying of Eq. (\ref{sech1}) means that the Universe has
experienced non-super-acceleration for a flat universe. From the above discussion,
it is clear that these energy conditions can be used to show
the evidence of acceleration and super-acceleration in the luminosity distance
redshift diagram. We consider a flat universe first. The luminosity distance is
\begin{equation}
\label{lum}
d_L(z)=(1+z)\int_0^z\frac{dz'}{H(z')}.
\end{equation}
The extinction-corrected distance modulus $\mu(z)=5\log_{10}[d_L(z)/{\rm Mpc}]+25$.
Substitute Eqs. (\ref{sech}) and (\ref{sech1}) into Eq. (\ref{lum}), we get
the upper bounds on the luminosity distance
\begin{equation}
\label{upperbound}
H_0 d_L(z)\le z(1+z), \quad H_0 d_L(z)\le (1+z)\ln (1+z).
\end{equation}
Again, Eq. (\ref{upperbound}) is derived from Eq. (\ref{sec}), but Eqs. (\ref{sec}) and
(\ref{upperbound}) are not equivalent because the luminosity distance involves integration.
To understand the integration effect, we use the
$\Lambda$CDM model as an example. For $\Lambda$CDM model, the Universe experienced accelerated expansion
in the redshift region $z\alt 0.76$ and decelerated expansion in the redshift region
$z\agt 0.76$. In other words, the strong energy condition is violated in the redshift
region $z\alt 0.76$, and satisfied in the redshift region $z\agt 0.76$.
We plot the distance modulus $\mu(z)$ for the $\Lambda$CDM model in Fig. \ref{lcdm}.
From Fig. \ref{lcdm}, we see that $\mu(z)$ for the $\Lambda$CDM model
is outside the bound given by the lower solid line up to the redshift $z\sim 7$. Of course, this does
mean that we see the evidence of acceleration for the $\Lambda$CDM model in the high
redshift region $z \simeq 7$.
In particular, from  this graph it is incorrect to conclude that  the strong energy
condition was first violated billions of years ago, at $z \ge 1$ in \cite{alcaniz}.
For more detailed discussions on the integration effects, see Ref. \cite{ygaz}.
What we can conclude from this graph   are: (a) The strong energy condition
Eq. (\ref{sec}) leads to the upper bound Eq. (\ref{upperbound})
on the luminosity distance, and the violation of Eq. (\ref{sec}) leads to the
violation of Eq. (\ref{upperbound}). (b) The satisfying of the upper bound Eq. (\ref{upperbound})
on the luminosity distance does not necessarily mean the satisfying of the strong energy
condition, and the violation of Eq. (\ref{upperbound}) does not mean the violation of the strong
energy condition. (c) The violation of Eq. (\ref{upperbound}) implies that the strong energy condition
was once violated, but not always violated. (d) If the upper bound Eq. (\ref{upperbound}) is satisfied, then
the strong energy condition Eq. (\ref{sec}) was once satisfied, but not necessarily always satisfied.

\begin{figure}
\centering
\includegraphics[width=12cm]{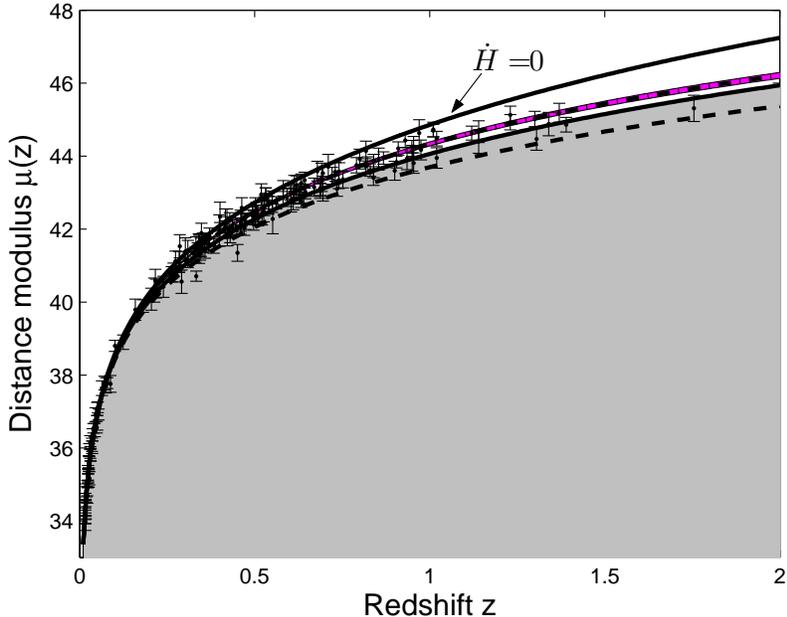}
\caption{The distance modulus $\mu(z)$. The solid lines denote the bounds
from the energy conditions. The dashed line corresponds to the model $\Omega_m=1$
and the dash-dotted line is for the $\Lambda$CDM model.}
\label{fig1}
\end{figure}

\begin{figure}
\centering
\includegraphics[width=12cm]{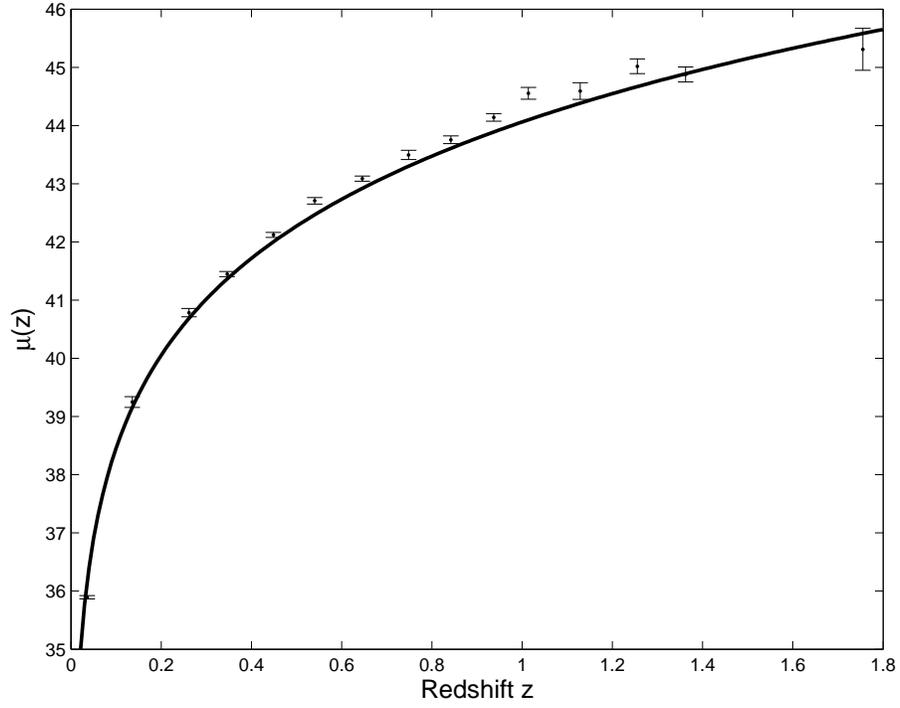}
\caption{The distance modulus $\mu(z)$. The solid line corresponds to $q(z)=0$.
The SN Ia data is the binned ESSENCE data with $1\sigma$ error.}
\label{fig1a}
\end{figure}

\begin{figure}
\centering
\includegraphics[width=12cm]{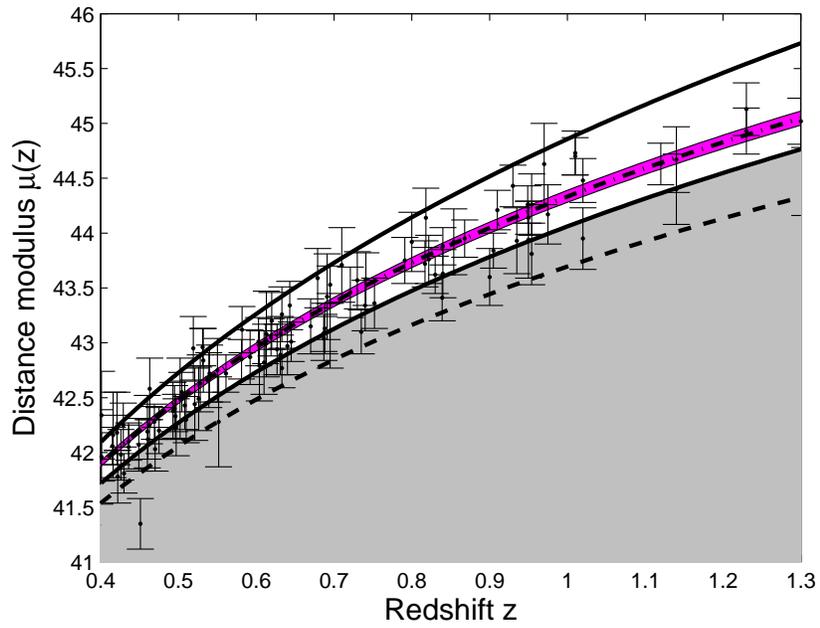}
\caption{The distance modulus $\mu(z)$ in the redshift region 0.4-1.3.}
\label{fig2}
\end{figure}

Now we are ready to apply the upper bounds (\ref{upperbound}) to the discussion of the acceleration
of the Universe. We plot these upper bounds on $\mu(z)$ in Fig. \ref{fig1}.
The lower solid line corresponds
to $q(z)=0$ and the upper solid line corresponds to $\dot{H}=0$.
If the Universe always experiences decelerated expansion, then the distance modulus
always stays in the shaded region. If some or all SN Ia
data are outside the shaded region, it means the Universe has accelerated once in
the past. On the other hand, if the Universe always experiences accelerated expansion,
then the distance modulus always stays above the lower solid line.
If all the SN Ia data are inside
the shaded region, it means that the Universe has once experienced decelerated expansion, but
it does not mean that the Universe has never accelerated.
Therefore, we can see the evidence of acceleration from the distance modulus graph directly
without invoking any cosmological model or any statistical analysis.
The ESSENCE SN Ia data \cite{riess06} is used to show the evidence of acceleration.
In Fig. \ref{fig1}, we show all the ESSENCE SN Ia data with $1\sigma$ error bars. The binned
ESSENCE SN Ia data is shown in Fig. \ref{fig1a}. In Fig. \ref{fig2}, we re-plot Fig. \ref{fig1}
in the redshift range $0.4\le z\le 1.3$.  From
Figs. \ref{fig1}- \ref{fig2}, it is evident that the Universe had accelerated in the past
because there are substantial numbers of SN Ia lying outside the shaded region.
We would like to stress that this conclusion is totally model independent.
There is no model or parametrization
involved in this conclusion. The assumptions we use are Einstein's general relativity and
the Robertson-Walker metric. For comparison, we also show the
model $\Omega_m=1$ (the dashed line) and the $\Lambda$CDM model with
$\Omega_m=0.27\pm 0.04$ (the dash dotted line); the $1\sigma$ error
is shown in the shaded region around the dashed dotted line. Note that due to integration effect,
even if some high $z$ SN Ia data are outside the shaded region, it does not mean that
we see evidence of acceleration in the high $z$ region. It is wrong to conclude that
the strong energy condition is violated in the high redshift region $z>1$. The correct
conclusion is that the strong energy condition was once violated in the past.

We may wonder about the low $z$ data. From Fig. \ref{fig1}, we see that $\mu(z)$ is almost
independent of any model. In fact, to the lowest order, $d_L(z)=H_0 z$. Because
the data are given with arbitrary distance normalization, so $H_0$ for this data
can be determined from the nearby supernova data $z\le 0.1$ with $d_L(z)=H_0 z$.
We find that $H_0=64.04$ and this value is used. Because $\mu(z)$ is almost
degenerate in the low redshift region $z\le 0.1$,
a current property of dark energy like $w_0$ is not well determined
from the SN Ia data. This was discussed in \cite{gong07,astier01,huterer,weller,alam,gong04,gong05}
with the help of the sweet spot of $w(z)$.

\section{Properties of dark energy}

\begin{figure}
\centering
\includegraphics[width=12cm]{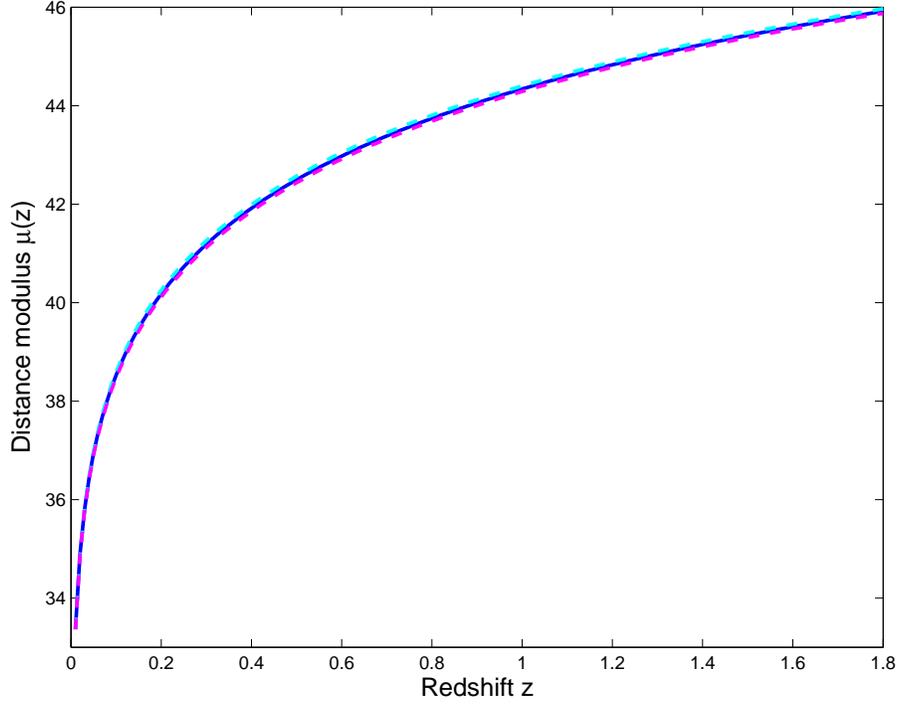}
\caption{The distance modulus $\mu(z)$ with $\Omega_{m}=0.27$, $w_0=-1.12$ and $w_a=0.59$.
The solid line is for the model (\ref{lind}).
The dashed lines are for the
model (\ref{wzeq}) with $w_1=-0.4$ and $w_1=-2$.}
 \label{fig5}
\end{figure}

In this section, we use a simple example to show that the choice of $w_0$ is not
a good one for exploring the property of dark energy. We use two models to show this.
The two models have the same behaviors at $z>z_c$ and different behaviors at $z\le z_c$,
where $z_c$ is arbitrary and small.
The first dark energy parametrization that we consider is \cite{lind}
\begin{equation}
\label{lind}
w(z)=w_0+\frac{w_a z}{1+z}.
\end{equation}
The dimensionless dark energy density is
\begin{equation}
\label{deneq}
\Omega_{DE}(z)=\Omega_{DE0}(1+z)^{3(1+w_0+w_a)}\exp[-3w_a z/(1+z)].
\end{equation}
By fitting this model to the essence data \cite{riess06}, we find that
$\chi^2=195.07$, $\Omega_m=0.35^{+0.14}_{-0.35}$, $w_0=-1.11^{+1.45}_{-0.93}$ and $w_a=-1.17^{+4.61}_{-17.48}$.
If we fix $\Omega_{m}=0.27$, then the best fit results are $\chi^2=195.17$, $w_0=-1.12\pm 0.44$
and $w_a=0.59^{+2.32}_{-2.54}$. From the Fisher matrix estimation, we find the sweet
spot is around $z=0.21$. If we fix $\Omega_{m}=0.27$ and $w_a=0.59$, then
$w_0=-1.12^{+0.09}_{-0.10}$ (1$\sigma$) $^{+0.17}_{-0.20}$ ($2\sigma$) $^{+0.25}_{-0.31}$ ($3\sigma$).
We plot the distance modulus for this model with $\Omega_{m}=0.27$, $w_0=-1.12$ and $w_a=0.59$
in Fig. \ref{fig5}.

\begin{figure}
\centering
\includegraphics[width=12cm]{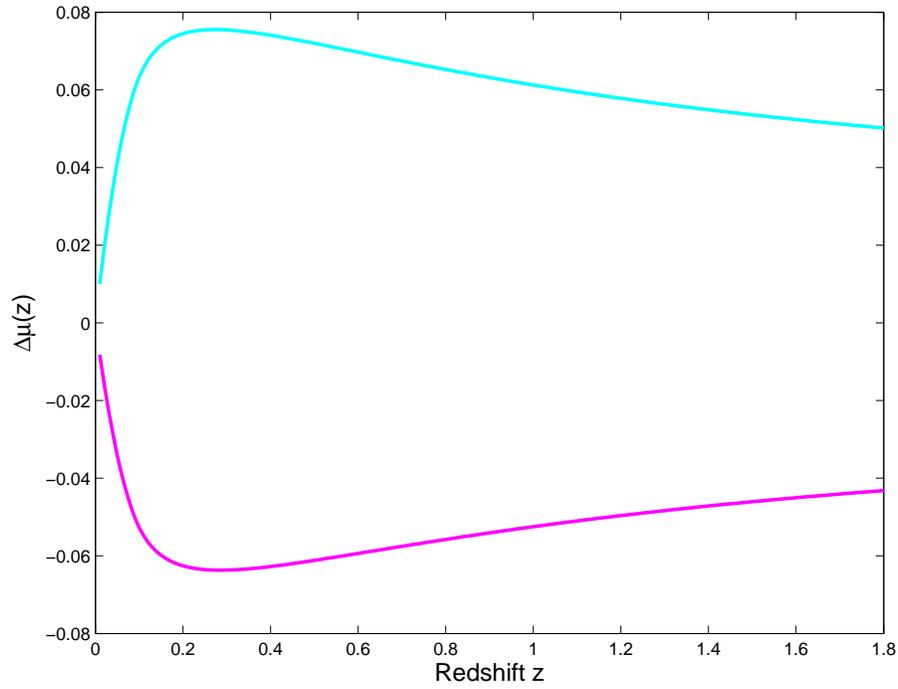}
\caption{$\Delta \mu(z)$ between the models (\ref{lind}) and (\ref{wzeq}).
The upper curve is for $w_1=-2$ and the lower curve is for $w_1=-0.4$.}
\label{fig6}
\end{figure}

\begin{figure}
\centering
\includegraphics[width=12cm]{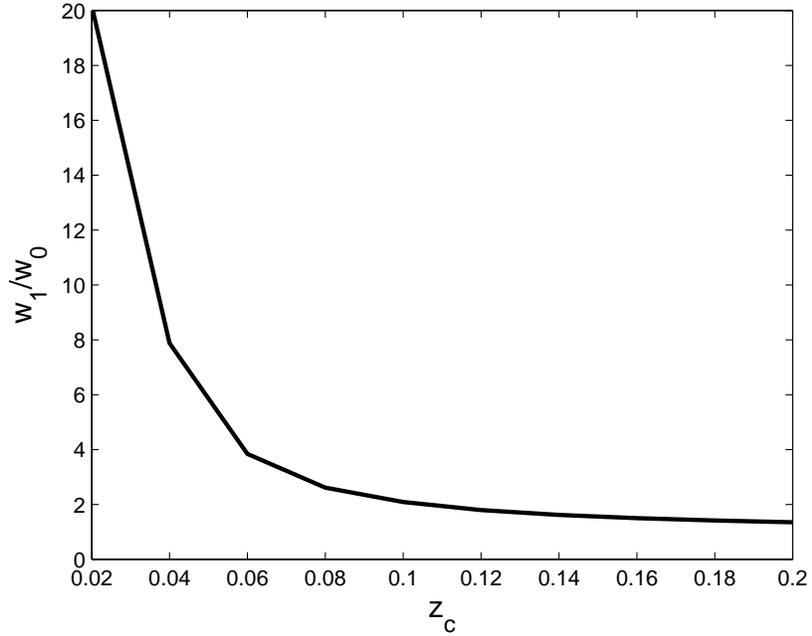}
\caption{The 2$\sigma$ lower value of $w_1$ normalized by the 2$\sigma$ lower value of $w_0$
as a function of $z_c$.}
\label{fig7}
\end{figure}

The second dark energy parametrization that we consider is
\begin{equation}
\label{wzeq}
w(z)=\begin{cases}
w_1+w_2 z,& z\le z_c,\\
w_0+w_a z/(1+z), & z>z_c,
\end{cases}
\end{equation}
where $w_2=(w_0-w_1)/z_c+w_a/(1+z_c)$.
The dimensionless dark energy density is
\begin{equation}
\label{deneq1}
\Omega_{DE}(z)=\begin{cases}
\Omega_{DE0}(1+z)^{3(1+w_1-w_2)}\exp(3w_2 z),& z\le z_c,\\
\Omega_{DEi}(1+z)^{3(1+w_0+w_a)}\exp[-3w_a z/(1+z)],& z>z_c,
\end{cases}
\end{equation}
where $\Omega_{DEi}=\Omega_{DE0} (1+z_c)^{3(w_1-w_2-w_0-w_a)}$ $\exp[3z_c(w_2+w_a/(1+z_c))]$.
We choose $\Omega_{m}=0.27$, $w_0=-1.12$, $w_a=0.59$ and $z_c=0.1$. If we take $w_1=-2.5$, we get
$\chi^2=197.90$. If $w_1=-2.0$, $\chi^2=196.12$. If $w_1=-1.5$, $\chi^2=195.21$.
If $w_1=-0.5$, $\chi^2=196.49$. If $w_1=-0.4$, $\chi^2=196.88$. In the first model (\ref{lind}),
we find $-1.43\le w_0\le -0.85$ at the $3\sigma$ level. However,
$w_1=-0.4$ and $w_1=-2.0$ are just a little more than $1\sigma$ away from $w_0=-1.12$.
So we conclude that $w(z=0)$ is not well constrained by the SN Ia data.
The distance modulus for this model with $w_1=-0.4$ and $w_1=-2.0$ are shown in Fig. \ref{fig5}.
In Fig. \ref{fig6}, we plot the differences between the model (\ref{wzeq}) and the model (\ref{lind})
for $w_1=-0.4$ and $w_1=-2.0$. For completeness, we also vary $z_c$ in the model (\ref{wzeq}) and
compare the 2$\sigma$ error of $w_1$ with that of $w_0$ for different choice of $z_c$. We plot
the result in Fig. \ref{fig7}. For bigger $z_c$, the difference becomes smaller which is consistent
with the appearance of the sweet spot around $z=0.21$. At lower redshift, the models become more
degenerate and the error bar of $w(z)$ becomes bigger.

\begin{figure}
\centering
\includegraphics[width=12cm]{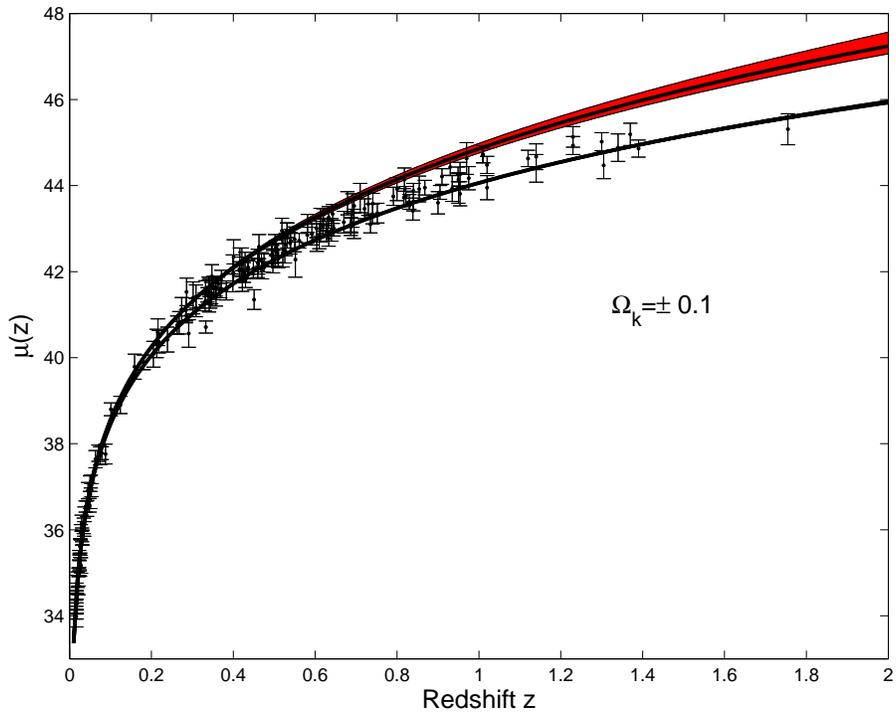}
\caption{The upper bounds on $\mu(z)$ from the energy conditions. The solids line correspond to
the flat universe $\Omega_k=0$. The shaded regions around these two lines correspond to
$\Omega_k=\pm 0.1$.}
\label{fig8}
\end{figure}

\section{Non-flat Universe}

When $k\neq 0$,
the luminosity distance becomes
\begin{eqnarray}
\label{lumdis}
d_{\rm L}(z)=\frac{1+z}{H_0\sqrt{|\Omega_{k}|}} {\rm
sinn}\left[\sqrt{|\Omega_{k}|}\int_0^z
\frac{dz'}{E(z')}\right],
\end{eqnarray}
where ${\rm sinn}(\sqrt{|k|}x)/\sqrt{|k|}=\sin(x)$, $x$, $\sinh(x)$ if $k=1$, 0, $-1$.
Substitute the lower bounds on $H(z)$ in Eqs. (\ref{sech}) and (\ref{sech1}) into Eq. (\ref{lumdis}), we
get upper bounds on $d_L(z)$. We plot the upper bounds for $\mu(z)$ in Fig. \ref{fig8}.
In this plot, we choose $\Omega_k=\pm 0.1$ which satisfies the observational constraint.
From Fig. \ref{fig8}, we see that even with $\Omega_k$ as large as $\pm 0.1$, it is
evident that the Universe had experienced accelerated expansion.

\section{Discussion}
The energy conditions $\rho+3p\ge 0$ and $\rho+p\ge 0$ give lower
bounds (\ref{sech}) and (\ref{sech1}) on the Hubble parameter
$H(z)$, and upper bounds on the distance modulus. If some SN Ia data
are outside the region bounded by Eq. (\ref{sech}), then we conclude
that the Universe had experienced accelerated expansion. In other
words, the distance modulus-redshift graph can be used to provide
direct model independent evidence of accelerated and
super-accelerated expansion.  If some SN Ia data are outside the
region bounded by Eq. (\ref{sech1}), then we conclude that the
Universe experienced super-accelerated expansion for a flat
universe. Unlike the usual parametrization methods, there is no
statistical analysis involved; all we need to do is to put the SN Ia
data in the distance modulus-redshift graph and see if there are
substantial numbers of SN Ia data lying outside or inside the bound
given by the strong energy condition. However, this direct probe
does not provide us any detailed information about the acceleration,
nor the nature of dark energy. Because the luminosity distance is an
integral of the Hubble parameter, the distance modulus does not give
us any information about the transition from decelerated expansion
to accelerated expansion. As we see from the $\Lambda$CDM model in
Fig. \ref{lcdm}, even when the Universe was experiencing decelerated
expansion in the past, the distance modulus may still stay outside
the bounded region for a while. Due to the same reason, the distance
modulus may satisfy the lower bound when the Universe is
accelerating \cite{ygaz}. The interpretation of the bounds on the
distance modulus is very important. These bounds provide the
evidence of acceleration or deceleration only, and they gave no
information on how and when the acceleration happened.

At low redshift $z\le 0.1$, the distance modulus is almost the
same for all the cosmological models. For example, the
difference of the distance modulus at $z=0.1$ between the $\Omega_m=1$
and $\Omega_\Lambda=1$ models is 0.16. This is well within
the current observational limit. It is even difficult for future
nearby SN Ia observation to reach limit below this uncertainty
due to intrinsic systematics and peculiar velocity dispersion. Therefore, the property
of dark energy at low redshift cannot be well constrained. This point was
discussed in \cite{gong07,astier01,huterer,weller,alam,gong04,gong05}
with the help of the sweet spot of $w(z)$. We use
two dark energy models which are identical at $z>0.1$ and differ at $z\le 0.1$
to further support this argument. From the model (\ref{lind}),
we find that $-1.43\le w_0\le -0.85$ at the $3\sigma$ level.
At a little more than $1\sigma$ level, we get $-2.0\le w_0\le -0.4$
for the model (\ref{wzeq}). So the $w_0$-$w_a$ parameterizations do not provide definite
information about the nature of dark energy.

In conclusion, the energy conditions provide direct
and model independent evidence of the accelerated expansion. The bounds
on the distance modulus also provide some directions for the future
SN Ia observations. Unfortunately, the method has some serious limitations.
It does not provide any detailed information about the acceleration
and the nature of dark energy.

\begin{acknowledgments}
Y.G. Gong and A. Wang thank J. Alcaniz for valuable discussions.
Y.G. Gong is supported by NNSFC under grant No. 10447008 and 10605042,
SRF for ROCS, State Education Ministry
and CMEC under grant No. KJ060502.
A. Wang is partially supported by VPR funds, Baylor University.
Y.Z. Zhang's work is in part supported by NNSFC under
Grant No. 90403032 and also by National Basic Research Program of
China under Grant No. 2003CB716300.
\end{acknowledgments}



\begin{thebibliography}{nbound}
\bibitem{agr98} A.G. Riess  {\it et al.}, Astron. J. {\bf 116}, 1009 (1998);
S. Perlmutter {\it et al.}, Astrophy. J. {\bf 517}, 565 (1999).
\bibitem{riess} A.G. Riess {\it et al.}, Astrophys. J. {\bf 607}, 665 (2004).
\bibitem{astier} P. Astier {\it et al.}, Astron. and Astrophys. {\bf 447}, 31 (2006).
\bibitem{riess06} A.G. Riess {\it et al.}, astro-ph/0611572;
W.M. Wood-Vasey {\it et al.}, astro-ph/0701041; T.M. Davis {\it et al.}, astro-ph/0701510.
\bibitem{wmap3} D.N. Spergel {\it et al.}, Astrophys. J. Suppl. {\bf 170}, 377 (2007).
\bibitem{sdss} D.J. Eisenstein {\it et al.}, Astrophys. J. {\bf 633}, 560 (2005).
\bibitem{DE} V. Sahni  and A. A. Starobinsky, Int. J. Mod. Phys. D
{\bf 9}, 373 (2000); T. Padmanabhan, Phys. Rep.
{\bf 380}, 235 (2003); P.J.E. Peebles and B. Ratra, Rev. Mod. Phys. {\bf 75}, 559 (2003);
V. Sahni, {\it The Physics of the Early Universe}, edited by
E. Papantonopoulos (Springer, New York 2005), P. 141;
T. Padmanabhan, {\it Proc. of the 29th Int. Cosmic Ray Conf.}  10, 47 (2005);
E.J. Copeland, M. Sami and S. Tsujikawa, Int. J. Mod. Phys. D {\bf 15}, 1753 (2006).
\bibitem{virey} J.-M. Virey {\it et al.}, Phys. Rev. D {\bf 72}, 061302(R) (2005).
\bibitem{sturner} C.A. Shapiro and M.S. Turner, Astrophys. J. {\bf 649}, 563 (2006).
\bibitem{gong06} Y.G. Gong and A. Wang, Phys. Rev. D {\bf 73}, 083506 (2006).
\bibitem{gong07} Y.G. Gong and A. Wang, Phys. Rev. D {\bf 75}, 043520 (2007).
\bibitem{astier01} P. Astier, Phys. Lett. B {\bf 500}, 8 (2001).
\bibitem{huterer} D. Huterer and M.S. Turner, Phys. Rev. D {\bf 64},
123527 (2001).
\bibitem{weller} J. Weller  and A. Albrecht, Phys. Rev. D {\bf 65}, 103512 (2002).
\bibitem{alam} U. Alam, V. Sahni, T.D. Saini and A.A.
Starobinsky, Mon. Not. Roy. Astron. Soc. {\bf 354}, 275 (2004).
\bibitem{gong04} Y.G. Gong, Class. Quantum Grav. {\bf 22}, 2121 (2005).
\bibitem{gong05} Y.G. Gong and Y.Z. Zhang, Phys. Rev. D {\bf 72}, 043518 (2005).
\bibitem{lind} M. Chevallier and  D. Polarski, Int. J. Mod.
Phys. D {\bf 10}, 213 (2001); E.V. Linder, Phys. Rev. Lett. {\bf 90}, 091301
(2003).
\bibitem{jbp}  H.K. Jassal, J.S. Bagla and T. Padmanabhan,
Mon. Not. Roy. Astron. Soc. {\bf 356}, L11 (2005).
\bibitem{par1} T.R. Choudhury and  T. Padmanabhan, Astron. Astrophys. {\bf 429}, 807 (2005).
\bibitem{par2} J. Weller  and A. Albrecht, Phys. Rev. Lett. {\bf
86}, 1939 (2001); D. Huterer and G. Starkman, {\it ibid}. {\bf
90}, 031301 (2003).
\bibitem{par3} G. Efstathiou, Mon. Not. Roy. Astron. Soc. {\bf 310}, 842 (1999);
B.F. Gerke and G. Efstathiou, {\it ibid}. {\bf 335}, 33 (2002);
P.S. Corasaniti and E.J. Copeland, Phys. Rev. D {\bf 67}, 063521 (2003);
S. Lee, {\it ibid}. {\bf 71}, 123528 (2005);
K. Ichikawa and T. Takahashi, {\it ibid}. {\bf 73}, 083526 (2006);
J. Cosmol. Astropart. Phys. 02 (2007) 001;
C. Wetterich, Phys. Lett. B {\bf 594}, 17 (2004).
\bibitem{par4} U. Alam, V. Sahni and A.A.
Starobinsky, J. Cosmol. Astropart. Phys. 06 (2004) 008;
R.A. Daly and S.G. Djorgovski, Astrophys. J. {\bf 597}, 9
(2003); R.A. Daly and S.G. Djorgovski, {\it ibid}. {\bf 612}, 652 (2004).
\bibitem{par5} J. J\"{o}nsson, A. Goobar, R. Amanullah and L.
Bergstr\"{o}m, J. Cosmol. Astropart. Phys. 09 (2004) 007; Y. Wang and P. Mukherjee,
Astrophys. J. {\bf 606}, 654 (2004); Y. Wang and M. Tegmark, Phys.
Rev. Lett. {\bf 92}, 241302 (2004); V.F. Cardone, A. Troisi and S. Capozziello,
Phys. Rev. D {\bf 69}, 083517 (2004); D. Huterer and A. Cooray, {\it ibid}. {\bf 71}, 023506 (2005).
\bibitem{gong04a} Y.G. Gong, Int. J. Mod. Phys. D {\bf 14}, 599 (2005);
M. Szydlowski and W. Czaja, Phys. Rev. D {\bf 69}, 083507 (2004);
{\it ibid}. 083518 (2004); M. Szydlowski, Int. J. Mod. Phys. A {\bf 20}, 2443 (2005).
\bibitem{wang05} B. Wang, Y.G. Gong and R.-K. Su, Phys. Lett. B {\bf 605}, 9 (2005).
\bibitem{jbp05}  H.K. Jassal, J.S. Bagla and T. Padmanabhan, Phys.
Rev. D {\bf 72}, 103503 (2005).
\bibitem{nesseris6a} S. Nesseris and L. Perivolaropoulous,
J. Cosmol. Astropart. Phys. 01 (2007) 018; 02 (2007) 025.
\bibitem{berger} V. Berger, Y Gao and D. Marfatia, Phys. Lett. B {\bf 648}, 127 (2007).
\bibitem{sahni06} V. Sahni and A.A. Starobinsky, Int. J. Mod. Phys. D {\bf 15}, 2105 (2006);
U. Alam, V. Sahni and A.A. Starobinsky, J. Cosmol. Astropart. Phys. 02 (2007) 011.
\bibitem{lihong} H. Li {\it et al.}, astro-ph/0612060;
G.-Z. Zhao {\it et al.}, astro-ph/0612728; X. Zhang and F.-Q. Wu, Phys. Rev. D {\bf 76}, 023502 (2007);
L.-I. Xu, C.-W. Zhang, B.-R. Chang and H.-Y. Liu, astro-ph/0701519;
H. Wei, N.-N. Tang and S.N. Zhang, Phys. Rev. D {\bf 75}, 043009 (2007).
\bibitem{yun06} Y. Wang and P. Mukherjee, Astrophys. J. {\bf 650}, 1 (2006).
\bibitem{evldh} E.V. Linder and D. Huterer, Phys. Rev. D {\bf 67},
081303(R) (2003).
\bibitem{saini} T.D. Saini, Mon. Not. Roy. Astron. Soc. {\bf 344}, 129 (2003).
\bibitem{jbp06}  H.K. Jassal, J.S. Bagla and T. Padmanabhan,
astro-ph/0601389.
\bibitem{nesseris05} S. Nesseris and L. Perivolaropoulous, Phys.
Rev. D {\bf 72}, 123519 (2005);
P. Serra, A. Heavens and A. Melchiorri, astro-ph/0701338;
C. Zunckel and R. Trotta, astro-ph/0702695; A. Shafieloo, astro-ph/0703034.
\bibitem{visser} M. Visser, Phys. Rev. D {\bf 56}, 7578 (1997).
\bibitem{alcaniz} J. Santos, J.S. Alcaniz and M.J. Rebou\c{c}as, Phys. Rev. D {\bf 74}, 067301 (2006);
J. Santos, J.S. Alcaniz, N. Pires and M.J. Rebou\c{c}as, astro-ph/0702728, Phys. Rev. D {\bf 75}, 083523 (2007).
\bibitem{aasen} A.A. Sen and R.J. Scherrer, astro-ph/0703416.
\bibitem{ygaz} Y.G. Gong and A. Wang, arXiv: 0705.0996, Phys. Lett. B {\bf 652}, 63 (2007).
\end{thebibliography}
\end{document}